# Cellphone based Portable Bacteria Pre-Concentrating microfluidic Sensor and Impedance Sensing System

Authors: Jing Jiang[1#], Xinhao Wang[1#], Ran Chao[2,3], Yukun Ren[1], Chengpeng Hu[1], Zhida Xu[1] and Gang Logan Liu[1]*

[1]Department of Electrical and Computer Engineering, University of Illinois at Urbana-Champaign

[2]Department of Chemical and Biomolecular Engineering, University of Illinois at Urbana-Champaign

[3]Department of Bioengineering, Department of Chemistry, Center for Biophysics and Computational Biology and Institute for Genomic Biology

[#]These authors contributed equally

*Corresponding author. Tel. +1 217-244-4349; E-mail: loganliu@illinois.edu

## Abstract

Portable low-cost sensors and sensing systems for the identification and quantitative measurement of bacteria in field water are critical in preventing drinking water from being contaminated by bacteria. In this article, we reported the design, fabrication and testing of a low-cost, miniaturized and sensitive bacteria sensor based on electrical impedance spectroscopy method using a smartphone as the platform. Our design of microfluidics enabled the pre-concentration of the bacteria which lowered the detection limit to 10 bacterial cells per milliliter. We envision that our demonstrated smartphone-based sensing system will realize highly-sensitive and rapid in-field quantification of multiple species of bacteria and pathogens.

## Introduction

With the improving life quality in many developed and developing countries, citizens are more concerned about food and water safety[1] and eager to know whether their drinking water or food are contaminated by nitrate [2], harmful bacteria[3] or heavy metal ions[4]. Some species of *Escherichia Coli* (*E. coli*) and Salmonella bacteria have been known to have caused serious food and water contamination issues which can severely threaten civilians' health conditions [5]. In order to quantify the amount of bacteria quickly, researchers have developed fluorescence detection techniques and DNA-biosensors to count *E. coli* [6] for various bacteria detections. However, fluorescence detection needs dye labeling [7] which requires high expense and professional training for operation. DNA-biosensor is also required to perform Polymerase chain reaction (PCR) as the first step[8] which is a complicated procedure in need of complicated facilities. So researchers also invented a series of label-free detection methods including quartz crystal microbalance (QCM) [9], microfluidic [10] and electrochemical methods[3]. Some of the techniques can detect the bacteria rapidly and accurately, however most of these detection methods have to be performed in specialized laboratory environments with the assistance of sophisticated equipment. Among these detection techniques, the electrochemical impedance spectroscopy (EIS) method is able to elucidate the electronic and physical properties of electrochemical systems including diffusion coefficients, adsorption mechanisms, capacitances, charge transfer resistances, and electron transfer rate constants. Due to its sufficient sensitivity, simplicity and cost-effectiveness, it has been increasingly applied in bio-sensing with numerous methods in the past few years [11]. It has also been implemented as a label-free detection tool for quantification of bacteria with minimal sample preparations.[12] Additionally, by coating the electrodes with different antibodies, people used EIS method to detect different pathogens like *E. coli* O157 [13]. However, high limit of detection and low sensitivity have prevented the application of EIS from being used practically in-field as a bacteria sensor. Meanwhile, microfluidic chips based measurement platforms which are affordable, portable and accessible to the public[14] have not been developed yet. Fortunately, as the smartphone is becoming more popular in our daily life, researchers have started to explore for the possibilities to make use of this powerful and portable platform for biological sensing.

In 2012, smartphone is becoming more prevalent in the US market with 115.8 million users [15] accounting for ~37% of US population, expected to rise to ~61% by 2016[15]. With the integration of GPS[16], powerful CPU, touch-screen displays, internet connection capability, and intelligent operating system[17], smartphones are able to provide extensive user-friendly services. Affordable smartphone peripheral devices with sensing capabilities will immensely help citizens to learn more information in environment like air and water quality at anywhere and anytime[18,19]. Bacteria sensing information collected by the phones can also be transmitted to cloud computing sites through 3G/4G network for further data processing and establishing a participatory water-borne bacteria sensing map on the internet for information broadcasting.

In this paper, we report the design, fabrication and integration of a low-cost, hand-held, and sensitive microfluidic EIS bacteria pre-concentrator and sensor based on a smartphone through wireless connection. It enables smartphone users to detect the density of as low as 10 *E. coli* cells per milliliter of water. Our integrated microfluidic sensor has interdigitated sensing electrodes on micro-hole array silicon substrate and a sensing microfluidic chamber bounded by a nano-porous filter paper which is also used to pre-concentrate bacteria in sample solutions. A specifically-designed impedance network analyzer chip with a microcontroller together performs EIS measurement and analysis. An Android application program (App) has been developed to enable recording and visualization of testing results as well as control of the sensor electronics. The real-time measurement data will be transmitted to a smartphone by a Bluetooth circuit module.

**Methods**

*Principle and design*

The principle of our miniaturized bacteria sensor is EIS theory which has been developed and applied in bacteria quantification [13,20,21] for more than a decade. According to the Randles model, the equivalent circuit includes ohmic resistance ($R_s$) of electrolyte, Warburg impedance ($Z_w$) resulted by the diffusion of ions from bulk electrolyte to the interdigitated electrodes, electron-transfer resistance ($R_{et}$) and double layer capacitance ($C_{dl}$) shown in Fig. 1(a). $R_s$ and $Z_w$ represent the features of the electrolyte solution diffusion at the probe, while $C_{dl}$ and $R_{et}$ depend on the insulating and dielectric properties at the interface of electrolyte and electrodes and are affected by the property change occurring at the electrode interface. The distribution of bacterial cells between the interdigitated electrodes affects the interfacial electron-transfer kinetics thereby increase or decrease the electron-transfer conductivity in electrolyte environment. $R_{et}$, the electron transfer resistance is a parameter that can be observed at higher frequencies corresponding to the electron-transfer-limited process and can be measured as the diameter of the semicircle portion in the Nyquist plot in Fig. 1 (b). The intercept of the semicircle with the $Z_{re}$ or the real axis at high frequency is equal to $R_s$. The linear part in Fig. 1(b), which is the characteristic at lower frequencies, represents the diffusion-limited processes.

The design of multi-stage filtering is comprised of one layer of silicon chip having a large array of through holes with diameter of 10 microns and one layer of nano-porous filter paper with submicron pore size. Fig 1 (c) shows the cross-sectional view of the integrated bacteria sensor. When the sample solution containing bacteria is injected from the bottom inlet, large particles in the solution will be blocked by the silicon filter while bacteria of our interest can go through the 10-micron-in-diameter silicon holes and are then blocked by the nanoporous filter, staying in the microfluidic sensing chamber. The pre-concentrated bacterial cells distributed around the interdigitated electrodes are subject to impedance sensing. The entire integrated components are packaged by Polydimethylsiloxane (PDMS) material. Fig 1 (d) shows the concept design of our packaged device. Users can take a certain amount of suspicious water sample which may contain bacteria into a syringe, and then inject the liquid through the channel as shown in Fig. 2 (a). Since all the bacteria are blocked by the filter and only 30 μL of liquid remains as a result of our geometric design, this sensor package allows users to pre-concentrate the bacteria solution before the measurement. The capacities of most standard syringes range from 1 mL to 60 mL, so we can pre-concentrate the bacteria by 10 to 2000 folds after excess liquid leaves the chamber from the outlet. Thus, our detection limit can be improved to as low as 10 bacterial cell per milliliter. The actual miniaturized bacteria sensor package is as large as one US quarter-dollar coin as shown in Fig 2 (b).

*Wireless system for the sensor*

A schematic diagram illustrating the main components of the wireless system is shown in Fig 2 (c). Our wireless sensor system has been designed applicable for most of Bluetooth transceiver as well as for Android phones [22]. Our system includes an Android cellphone (HTC ONE X), a Bluetooth shield (Seeed SLD63030P), a micro controller (Arduino), a chip for impedance converter network analyzer (AD 5933), and our packaged sensor.

An Android App has been developed for users to set the start/end frequencies and the frequency sweeping step size in impedance analysis. "Connect" button allows the cellphone to connect the sensor through Bluetooth. The benefits of Bluetooth connectivity include efficient power consumption of less than 10 mW[23] and standardization for smartphones and computers. Then Arduino microprocessor board generates corresponding commands according to the input parameters from the smartphone and asks the AD 5933 chip to send out sinusoidal signals to the bacteria sensor. Depending on the concentration of the bacteria, the corresponding signals acquired by the AD 5933 chip is sent back to the smartphone through the Arduino board and the Bluetooth shield. After this process completes, the smartphone App plots the impedance value with respect to frequency on the screen as shown in Fig 2 (c). It can also calculate a calibration curve after measuring several standard bacterial solutions, and then users can measure the bacterial concentration of an unknown sample.

Fig 2 (d) illustrates the basic diagram of the wireless LCR sensing platform which communicates with the mobile phone through the Bluetooth and can drive our packaged bacteria sensor to quantify bacteria. Microcontroller writes commands to the impedance converter chip and passes the start\end frequency and sweeping step data from the cellphone to the chip. The on-chip oscillator module of impedance converter generates corresponding sinusoidal waves as the input signal of the bacteria sensor. The output signals of the bacteria sensor containing attenuated amplitude and phase change information are analyzed by an on-chip digital signal processor with 1024 points Discrete Fourier Transform. Real and imaginary parts of the results are sent to the micro-controller for converting into impedance and phase information. Eventually, these results are transmitted back to the cellphone through the Bluetooth and displayed on the smartphone screen. After calibration by standard solutions, our platform and app can provide accurate quantification information for tested bacterial solution.

*Sensor Microfabrication*

The sensing part of miniaturized bacteria sensor is a pair of interdigitated electrodes fabricated on a piece of silicon chip with micro-scale through-hole arrays. Because interdigitated microelectrodes have advantages over conventional electrodes for analytical measurements including high signal-to-noise ratio, low resistance, small solution volumes requirement and rapid attainment of steady state,[24] we adopt this design as the sensing part of our sensor in this research. The top view image of this part is shown in Fig 3 (a). The diameter of the holes is 16 μm. The spacing between the electrodes is 20 μm while the width of the electrodes is 10 μm. Starting from 380 μm thick silicon wafer (University wafer 1815, p type, Boron doped, 4" 5~10 Ohm), 1 μm of $SiO_2$ was deposited with Plasma-Enhanced Chemical Vapor Deposition system (PlasmaLab) first at 300 ℃. Then, a layer of 100 nm $Al_2O_3$ which would be used as a hard mask for through holes was deposited uniformly and firmly on the $SiO_2$ layer by Atomic Layer Deposition (ALD) system (Cambridge NanoTech) at 250 ℃ as what Fig 3 (b)① shows. $Al_2O_3$ deposited by ALD has been proved to be a good mask for fluoride based Si deep reactive ion etching (DRIE) in our experiment. 100 nm $Al_2O_3$ allows ~200 um Si to be etched through. Then, photolithography and patterning were done with AZ 5214 photoresist and Karl Suss aligner as shown in Fig 3 (b)②. 15 nm of Ti for adhesion and 200 nm of gold were deposited by electron beam evaporation (Temescal). After the metal on the photoresist was lifted off by acetone bath, the gold layer on the exposed region remained as the pair of interdigited electrodes as shown in Fig 3 (b)③. After the fabrication of electrodes, another layer of micro-hole array mask was aligned and patterned. The holes were located between the electrodes and were designed for the bacteria to pass through while blocking big dirt particles in water samples. Buffered oxide etcher (1:10 HF: $NH_4F$) was applied to etch the $Al_2O_3$ and $SiO_2$ (Fig 3 (b)④, Fig 3 (b)⑤) hard mask. Then, $XeF_2$ etching (XACTIX) was performed on the back side of the silicon on the exposed part to thin down the silicon by half of its total thickness so that the micro-holes could be etched through by Bosch process with STS Advanced Silicon Etcher (Fig 3 (b)⑥, Fig 3 (b)⑦) for only around 1.5 hrs.

**Result and discussion**

*Concentration Measurement*

*E. coli* samples were provided by Institute for Genomic Biology from University of Illinois at Urbana and Champaign with the strain number of DH5a. The *E. coli* stock solution was centrifuged and rinsed with deionized water twice and diluted by 10, 100 and 1000 times respectively. The *E. coli* concentration for the stock solution is about $10^7$ ml$^{-1}$ measured by hemocytometer.

The prepared *E. coli* solution was injected into our device in the order of increasing concentration. To test the performance of packaged sensor electronics, a commercial LCR meter, Agilent 4284A, was used as the benchmark to measure the bacteria impedance by sweeping the signal frequency from 1 KHz to 1 MHz with 1 KHz step size.

Figure 4 (a) shows the Nyquist plot of the results of bacteria impedance measurement. At low frequency the bacteria sensing is mass transfer controlled, while at high frequency the bacteria sensing is kinetics controlled. In the kinetics controlled region, the diameter of the semicircle on Nyquist plot indicates the electron transfer resistance $R_{et}$. We found the relationship between bacteria concentration and electron transfer resistance in logarithmic form in Fig 4 (b). The fitting formula is:

$$\log R_{et} = -0.153 \log C_{bac} + 4.187 \tag{1}$$

The electron transfer resistance $R_{et}$ and bacteria concentration $C_{bac}$ could be related by Randles equivalent circuit of electrochemical impedance spectroscopy. According to Fig 1 (a) and (b), at high frequency, the Warburg impedance becomes negligible compared with $R_{et}$. Therefore the Faradaic impedance could be simplified to only electron transfer resistance $R_{et}$. Due to charge-transfer kinetics, the electron transfer resistance could be defined as:

$$R_{et} = \frac{RT}{nFi_0} \tag{2}$$

Here R is gas constant; T is absolute temperature; n is the number of transfer electrons; F is Faraday constant and $i_0$ is exchange current. In the theory of electrode kinetics, the exchange current at equilibrium condition could be defined as

$$i_0 = FAk^0 C_O^{*1-\alpha} C_R^{*\alpha} \tag{3}$$

$k^0$ is standard heterogeneous rate constant. Since *E. coli* is negative charged in neutral pH environment [25], we can assume that electrons spread out on the surface of bacteria so that the movement of cells contribute to the electron transfer current. The concentration of bacteria could be expressed as

$$C_{bac} = C_O^{*\beta_1} = C_R^{*\beta_2} \tag{4}$$

The exchange current could be simplified as:

$$i_0 = FAk^0 C_{bac}^{\beta} \tag{5}$$

In this case, the relation between electron transfer resistance and bacteria concentration could be expressed as: $R_{et} \propto C_{bac}^{-\beta}$. Figure 4 (b) displays this relationship from experimental data and parameter $\beta$ equals 0.153, which is consistent with theoretical derivation. Furthermore, in the condition of high frequency, the semicircle in Nyquist plot could be expressed as the function of real part ($Z_{Re}$) and imaginary part ($Z_{Im}$) as:

$$(Z_{Re} - R_s - \frac{R_{et}}{2})^2 + Z_{Im}^2 = (\frac{R_{et}}{2})^2 \quad (6)$$

The imaginary peak point on the semicircle satisfies the relationship:

$$\omega = 1/R_{ct}C_{dl} \quad (7)$$

$\omega$ is the voltage frequency while $C_{dl}$ is the double layer capacitance. By considering the Gouy-Chapman double layer model, which involves a diffusion layer of charge in the solution, the double layer capacitance can be related to bacteria concentration in logarithmic form as linearity. This relationship could be proved by the fitting curve in Fig 4 (b).

Here is one issue worth noting. In most other journal articles for bacterial detection [21], the electron-transfer resistance ($R_{et}$) increases as the concentration of bacteria increases. However, our testing data show that $R_{et}$ decreases as the bacteria concentration increases. To explain this phenomenon, we need to review Fig. 1 (a) where $R_s$ is the ohmic resistance of the electrolyte. In other cases, researchers used conductive electrolyte like 1x Phosphate-buffered saline (PBS) whose high conductivity is associated with low $R_s$. In those models [21], antibody treated electrodes trapped bacterial cells close to the electrodes and the double layer of lipid bilayer membrane of the cells retarded the electron transfer in the electrolyte. As a result, $R_{et}$ increased while the concentration of cells increased. In our case, deionized (DI) water (18 MOhm) was applied to dilute *E. coli* bacteria after centrifuging and rinsing from culture medium before testing, since *E. coli* at around neutral pH environment carries charge[25], higher concentration of *E. coli* induced resistance $R_{et}$ lower than the resistance of electrolyte, $R_s$.

The reason why we used DI water as the background solution to detect the concentration of bacteria is that our project investigated a new method for direct detection and quantification of bacteria on mobile detection platform in no need of pretreating field water samples, like centrifuging and diluting with 1x PBS and other redox molecule additives like $[Fe(CN)_6]^{3-/4-}$. Current result is a proof of concept to find the correlation between bacteria concentration and EIS in natural waters.

*Smartphone software application*

The software was developed on Android operating system. Implemented with multiple functions of sending sensor control commands, receiving data and plotting, this software is able to remotely control microcontroller (Arduino) through a Bluetooth Shield transceiver board. Connected with the miniaturized bacteria sensor, the smartphone App is able to implement the same functions as a commercial LCR meter does. The user interface of the App is shown in Fig 2(c).

In order to characterize the sensing capability of the smartphone bacteria sensor system, we have performed the tests with different concentrations of *E. coli* solutions and measured impedance spectra by sweeping the frequency from 2 kHz to 100 kHz, as shown in Fig. 5 (a). Figure 5 (b) and (c) show the magnified chart of Fig. 5 (a). We diluted the stock bacteria solution into different concentrations and used a 60 ml syringe to let 60 ml of calibration sample solutions with the cell concentrations of 10 ml$^{-1}$, 100 ml$^{-1}$, 1,000 ml$^{-1}$, and 10,000 ml$^{-1}$ to pass through the sensor. A Nyquist diagram was plotted on the smartphone screen as the thin lines in Fig 5 (b). Then the program can automatically extract the $R_{et}$ value by calculating the peak imaginary resistance from the curve in Fig 5 (b) and plot the $R_{et}$ versus *E. coli* concentrations diagram as shown in Fig 5 (c). A blue fitting curve has been derived as, where x is the concentration of *E. coli* and y is the imaginary impedance peak, $R_{et}$.

$$y = 79249x^{-0.21} \tag{8}$$

Then we performed a measurement of prepared bacteria solution with calculated concentration of 333 cells per milliliter. The result is shown as the thick orange curve in Fig 5 (b), where $R_{et}$ equals 25727.4 Ohm which is plotted as the green cross in Fig 5 (c). According to the fitting curve, the measured concentration is 212 cells per milliliter so that the relative error is 36.4% with respect to the actual concentration. Note that this tested cell concentration is extremely low with less than one cell per microliter.

One difference between the Nyquist plot results obtained by the bench-top LCR meter and those acquired by our wireless impedance sensing platform is the shape of the curve. First, the linear part disappeared in our sensor. Because the linear part represents the diffusion-limited process and corresponds to low-frequency response. While the bench-top LCR meter can sweep frequency starting from 1 kHz, our wireless impedance sensor starts sweeping from 2 kHz where kinetic control dominates. Second, since the upper frequency limit of the bench-top LCR meter is 1 MHz and that of our wireless sensing platform is only 100 kHz, the measurement results in the Nyquist plot converge to the origin for the bench-top equipment measurement but not for the wireless sensing platform. When we use $R_{et}$ for calibration, because $R_{et}$ peak for concentrations of bacteria higher than 1,000 cells/ml will fall out of the measurable frequency range, this instrument limitation shrinks the upper dynamic range to 1,000 cells/ml. As the $R_{et}$ peak exists for low concentration solution, the limit of detection is not degraded.

**Conclusions**

In summary, we have performed the design, fabrication and testing of a low-cost, miniaturized and sensitive wireless bacterial sensor that can pre-concentrate bacteria solution to obtain a detection limit of as low as 10 cells per milliliter. In order to enable citizens to perform EIS measurement conveniently and understand whether their drinking water has been contaminated, we designed and tested a smartphone-based miniaturized impedance spectroscopic measurement platform with Bluetooth connectivity. We have used commercial bench-top LCR meter to benchmark the performance and stability of our bacteria sensor. Additionally we integrated our sensor with the wireless impedance sensor platform to conduct a natural water sample testing after calibration. The limit of detection for the bacteria sensing is 10 *E. coli* cells per milliliter and its dynamic range is from 10 *E. coli* cells/ml to 1,000 cells/ml. We compared the measured *E. coli* concentration with the actual cell concentration, and got the result on the same order of magnitude with an error of 36.4%. Finally, we have also proved that our Android app in the smartphone worked properly with our low-cost wireless impedance bacteria sensing platform, which enables smartphone users to measure bacterial contamination in their daily-used water conveniently and cost-effectively. Moreover, the same wireless sensing platform and multi-stage pre-concentration filtering sensor package can be extended for specific pathogen detection by coating antibodies on the electrodes where users can detect the concentration of each kind of bacteria with a single device after performing one test.

**Acknowledgements**

This work was supported by the U.S. Army. We also thank Wenchuan Wei from Tsinghua University for helping debug the code to control AD5933 chip.

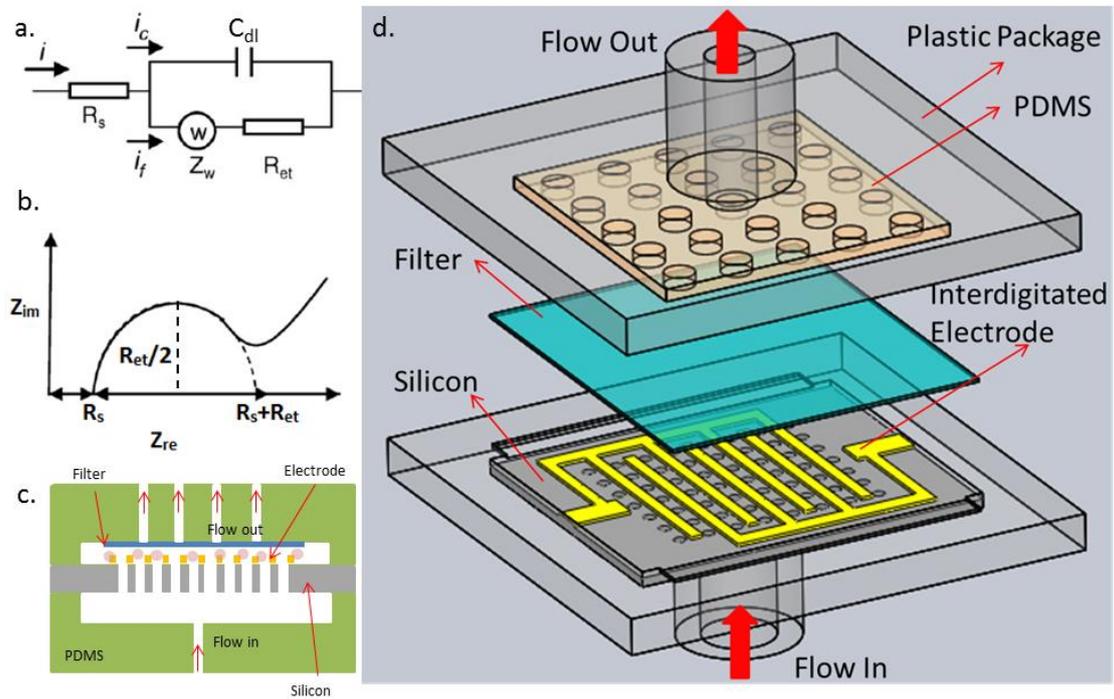

**Fig. 1** Electrochemical Impedance Spectroscopic (EIS) sensing principle. a. Randles equivalent circuit of EIS; b. a typical Nyquist plot for EIS; c. Cross-sectional view of the integrated EIS bacteria sensor. Bacteria will pass through micro-hole silicon filter and be blocked by nonporous filter above the interdigitated sensing electrodes; d. 3D model of the EIS bacteria sensor package.

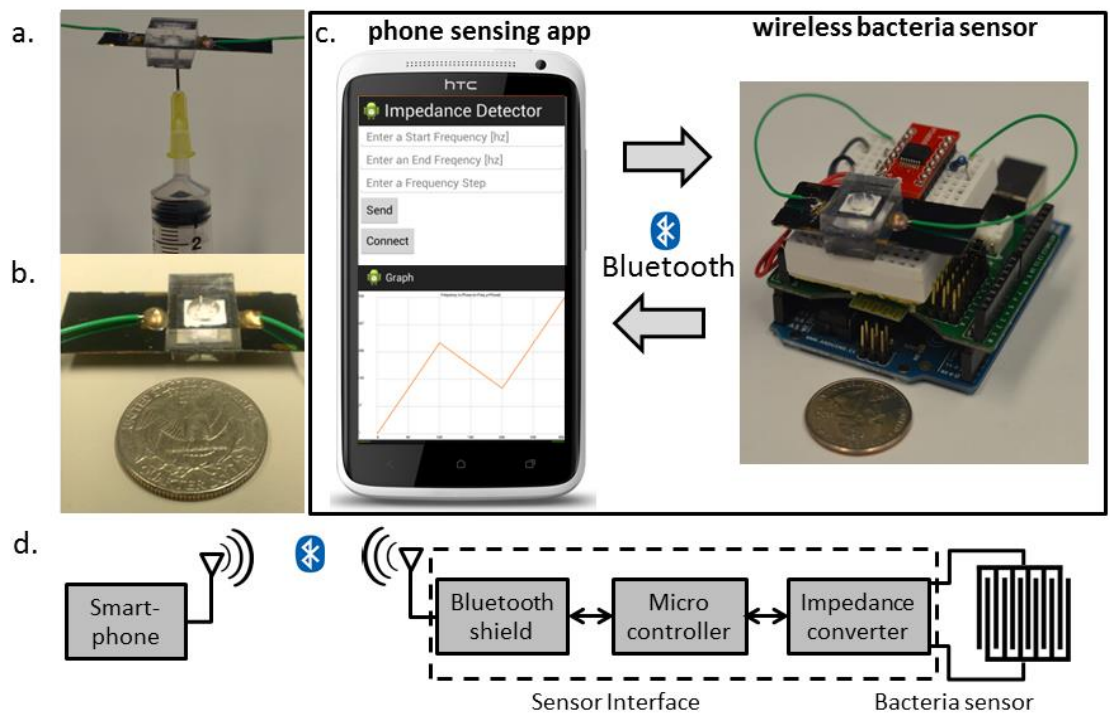

**Fig. 2** Wireless mobile phone bacteria sensing system. a. Picture showing syringe injection of testing liquid into the sensor package; b. Close view of the EIS bacteria sensor package; c. Picture showing communication scheme between smartphone sensing app and wireless bacteria sensor; d. Diagram of wireless sensing system.

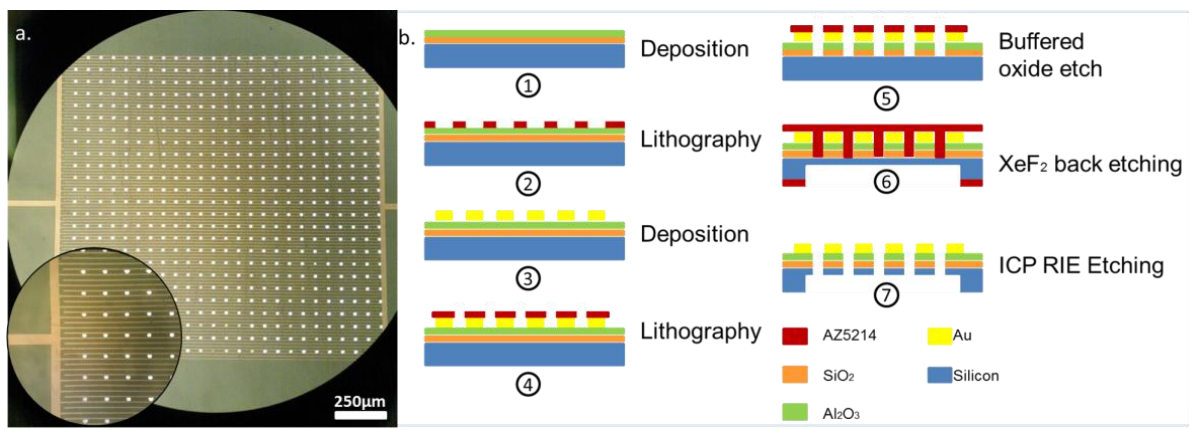

**Fig. 3** Sensor micro fabrication a. Top view microscopy images of the micro-hole array and interdigitated electrodes; b. Fabrication process of the silicon sensor chip.

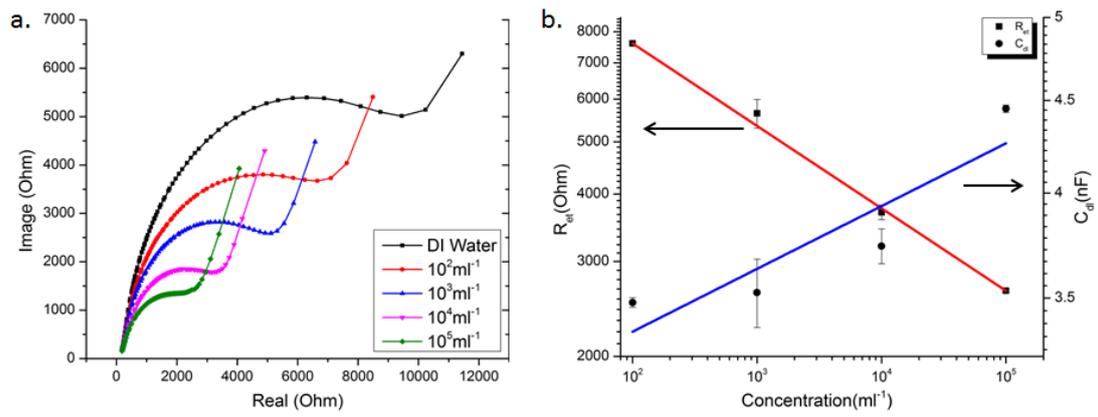

**Fig. 4** Two methods of concentration measurement using EIS bacteria sensor. a. Nyquist plots (real and imaginary parts of the complex impedance) of bacteria solutions at different concentrations; b. Logarithmic calibration between electron transfer resistance ($R_{et}$) and concentration and fitting curve (red line) and logarithmic calibration and fitting curve between double layer capacitance ($C_{dl}$) and concentration (blue line).

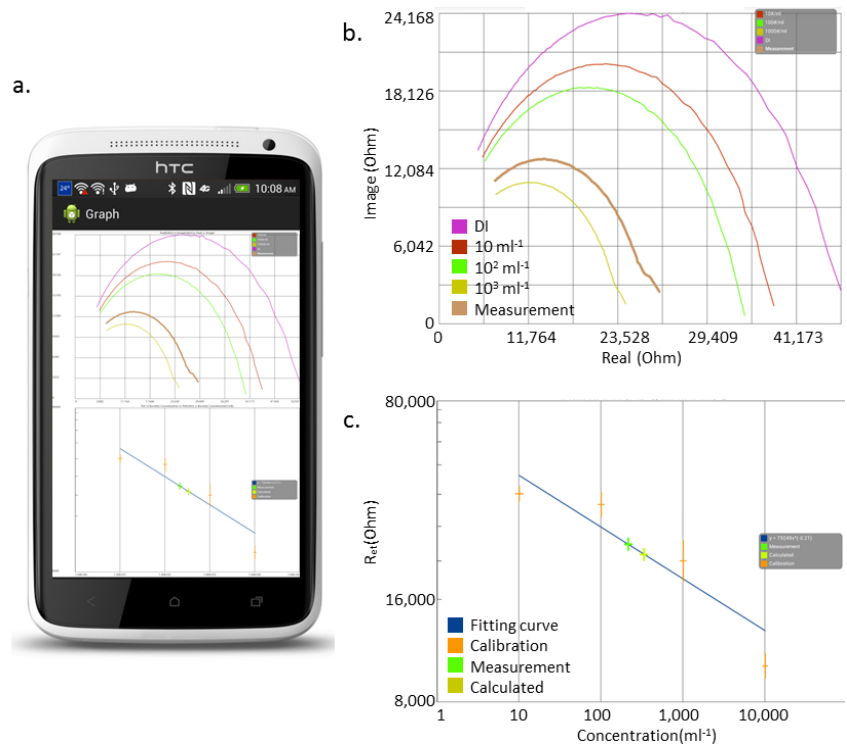

**Fig. 5** Calibration and sample measurement by the wireless cell phone bacteria sensing system a. Cellphone display of the impedance measurement results; b. Zoomed-in image of Nyquist plots on the cellphone, with four solutions of known bacteria concentrations used for sensor calibration, DI water and one sample solution to be tested; c. $R_{et}$ vs concentration plot and the fitted calibration curve. The *E. coli* concentration of the unknown sample was derived from fitting the measurement results into the calibration curve and corresponding formula, as the green cross shows; while the yellow shows the calculated (actual) concentration and corresponding $R_{et}$.

References

[1] V. Murray, G. Lau, Drinking Water Safety: Guidance to Health and Water Professionals—and other Health Protection Issues on Water Safety, (2011).

[2] M.R. Gartia, B. Braunschweig, T. Chang, P. Moinzadeh, B.S. Minsker, G. Agha, A. Wieckowski, L.L. Keefer, G.L. Liu, The microelectronic wireless nitrate sensor network for environmental water monitoring, J. Environ. Monit. (2012) 3068-3075.

[3] L. Wang, Q. Liu, Z. Hu, Y. Zhang, C. Wu, M. Yang, P. Wang, A novel electrochemical biosensor based on dynamic polymerase-extending hybridization for E. coli O157: H7 DNA detection, Talanta. 78 (2009) 647-652.

[4] Q. Liu, H. Cai, Y. Xu, L. Xiao, M. Yang, P. Wang, Detection of heavy metal toxicity using cardiac cell-based biosensor, Biosensors and Bioelectronics. 22 (2007) 3224-3229.

[5] N.J. Ashbolt, Microbial contamination of drinking water and disease outcomes in developing regions, Toxicology. 198 (2004) 229-238.

[6] X. Xue, J. Pan, H. Xie, J. Wang, S. Zhang, Fluorescence detection of total count of Escherichia coli and Staphylococcus aureus on water-soluble CdSe quantum dots coupled with bacteria, Talanta. 77 (2009) 1808-1813.

[7] K.M. Horsman, J.M. Bienvenue, K.R. Blasier, J.P. Landers, Forensic DNA analysis on microfluidic devices: a review, J. Forensic Sci. 52 (2007) 784-799.

[8] A. Battaglia, A.J. Schweighardt, M.M. Wallace, Pathogen Detection Using a Liquid Array Technology, J. Forensic Sci. 56 (2011) 760-765.

[9] J. Cai, C. Yao, J. Xia, J. Wang, M. Chen, J. Huang, K. Chang, C. Liu, H. Pan, W. Fu, Rapid parallelized and quantitative analysis of five pathogenic bacteria by ITS hybridization using QCM biosensor, Sensors Actuators B: Chem. 155 (2011) 500-504.

[10] Z. Wang, T. Han, T. Jeon, S. Park, S.M. Kim, Rapid detection and quantification of bacteria using an integrated micro/nanofluidic device, Sensors Actuators B: Chem. 178 (2013) 683-688.

[11] E.P. Randviir, C.E. Banks, Electrochemical impedance spectroscopy: an overview of bioanalytical applications, Analytical Methods. 5 (2013) 1098-1115.

[12] B. Chang, S. Park, Electrochemical impedance spectroscopy, Annual Review of Analytical Chemistry. 3 (2010) 207-229.

[13] L. Yang, R. Bashir, Electrical/electrochemical impedance for rapid detection of foodborne pathogenic bacteria, Biotechnol. Adv. 26 (2008) 135-150.

[14] D. Mark, F. von Stetten, R. Zengerle, Microfluidic Apps for off-the-shelf instruments, Lab on a Chip. 12 (2012) 2464-2468.

[15] Number of smartphone users in the U.S. from 2010 to 2016, (2012). www.statista.com

[16] J. Paek, J. Kim, R. Govindan, Energy-efficient rate-adaptive GPS-based positioning for smartphones, (2010) 299-314.

[17] H. Falaki, R. Mahajan, S. Kandula, D. Lymberopoulos, R. Govindan, D. Estrin, Diversity in smartphone usage, (2010) 179-194.

[18] N. Ramanathan, M. Lukac, T. Ahmed, A. Kar, P. Praveen, T. Honles, I. Leong, I. Rehman, J. Schauer, V. Ramanathan, A cellphone based system for large-scale monitoring of black carbon, Atmos. Environ. 45 (2011) 4481-4487.

[19] J. Li, Nanosensors-cellphone integration for extended chemical sensing network, Private communication to R Potyrailo. (2012).

[20] X. Yuan, C. Song, H. Wang, J. Zhang, Impedance and its Corresponding Electrochemical Processes, in: Anonymous Electrochemical Impedance Spectroscopy in PEM Fuel Cells, Springer, 2010, pp. 95.

[21] C. Ruan, L. Yang, Y. Li, Immunobiosensor Chips for Detection of Escherichia c oli O157: H7 Using Electrochemical Impedance Spectroscopy, Anal. Chem. 74 (2002) 4814-4820.

[22] J. Koetsier, Android captured almost 70% global smartphone market share in 2012, Apple just under 20%, (2013). venturebeat.com

[23] E. Ferro, F. Potorti, Bluetooth and Wi-Fi wireless protocols: a survey and a comparison, Wireless Communications, IEEE. 12 (2005) 12-26.

[24] K. Stulìk, C. Amatore, K. Holub, V. Marecek, W. Kutner, Microelectrodes. Definitions, characterization, and applications (Technical report), Pure and applied chemistry. 72 (2000) 1483-1492.

[25] D.A. Lytle, E.W. Rice, C.H. Johnson, K.R. Fox, Electrophoretic mobilities of Escherichia coli O157: H7 and wild-type Escherichia colistrains, Appl. Environ. Microbiol. 65 (1999) 3222-3225.